\documentclass[aps,nofootinbib,superscriptaddress]{JHEP3}
\usepackage{amsmath}

\def\be{\begin{equation}}
\def\ee{\end{equation}}
\def\ba{\begin{eqnarray}}
\def\ea{\end{eqnarray}}
\def\del{\partial}

\title{Cleaning up the cosmological constant}
\author{Ian Kimpton and Antonio Padilla
\\School of Physics and Astronomy, 
University of Nottingham, Nottingham NG7 2RD, UK} 

\date{\today}

\abstract{We present a novel idea for screening the vacuum energy contribution to the overall value of the cosmological constant, thereby enabling us to choose the bare value of the vacuum curvature empirically, without any need to worry about the zero-point energy contributions of each particle. The trick is to couple matter to a metric that is really a composite of other fields, with the property that the square-root of its determinant is the integrand of a  topological invariant, and/or a total derivative. This ensures that the vacuum energy contribution to the Lagrangian is non-dynamical. We then give an explicit example of a theory with this property that is free from Ostrogradski ghosts, and  is consistent with solar system physics and cosmological tests.}

\begin{document}



\section{Introduction}
A plethora of cosmological observations ranging from supernova \cite{sn} to the cosmic microwave background  \cite{wmap} seem to suggest that the energy density of the Universe is mostly in the form of dark energy, whose equation of state is consistent with that of a cosmological constant. If we did not care about particle physics we could happily treat the value of the cosmological constant as empirical, set by observation to be $\Lambda \sim H_0^2 \sim (10^{-33} \text{eV})^2$. The trouble is we do care about particle physics, and in standard General Relativity (GR), the bare value of $\Lambda$ is renormalised by the zero-point energy contributions of each particle. This latter contribution is enormous. As Pauli colourfully observed, without a dramatic amount of fine-tuning against the bare value of the cosmological constant, the vacuum energy contribution of the electron would be enough to ensure that the  cosmological horizon ``would not even reach to the moon" \cite{pauli}. Heavier particles make this even worse, as do vacuum energy contributions from phase-transitions  in the early Universe (eg electro-weak, QCD).  Supersymmetry provides a natural cut-off  for the corresponding energy density, of the order $(\text{TeV})^4$, but this is still {\it sixty} orders of magnitude larger than the dark energy scale $M_{pl}^2 H_0^2 \sim ( \text{meV})^4$.

One approach to the cosmological constant problem has been to identify some  symmetry argument or adjustment mechanism that forces the net cosmological constant to vanish, with the non-vanishing observed value coming from small (possibly quantum) fluctuations. Typically, such approaches will fall foul of Weinberg's  famous ``no-go" theorem  \cite{nogo}, although there are ways around that, as evidenced by  the SLED proposal \cite{sled} and the cosmology of  {\it the Fab Four} \cite{fab4}. Other recent attempts to address the cosmological constant problem include \cite{klink, bauer, Zonetti, Shaw, naresh}. However, for the most part, advocates of the $\Lambda$CDM model are inclined to sweep the troublesome vacuum energy  under the carpet. Here we take another approach: we clean up the cosmological constant by coupling matter to gravity in such a way that  the vacuum energy contribution  is eliminated completely. Now we fix the vacuum curvature empirically from observation, confident that particle physics considerations no longer play any role in determining its value. We should stress that we do not offer a dynamical explanation as to  why the vacuum curvature should take the empirical value $\Lambda \sim H_0^2 \sim (10^{-33} \text{eV})^2$. What we can say is that this choice is stable against radiative corrections in the Standard Model sector, something which is emphatically not the case when gravity is described by General Relativity. As a result,  we have removed the ugliest and most significant element of the cosmological constant problem. 
\section{A novel way to screen the vacuum energy}
Our idea is actually rather simple. We argue that all matter is minimally coupled to a metric, $\tilde g_{ab}$, that is not a fundamental field, rather it is a composite of fundamental fields, $\tilde g_{ab}=\tilde g_{ab} (\phi_i, \partial \phi_i, \del \del \phi_i, \ldots )$\footnote{As in GR, we assume that at least some of the  fields do not admit a {\it  locally} conserved energy-momentum tensor. This ensures that  we evade the no-go theorems presented in \cite{ww}. }. Note that we denote those fields collectively as $\phi_i$, suppressing their tensor rank for brevity. One of the $\phi_i$ will be the fundamental metric, $g_{ab}$, which is dynamical and distinct from the physical metric $\tilde g_{ab}$. There may be other fields too, in other words $\{ \phi_i\}=\{g_{ab}, \text{scalar fields, vector fields, other tensor fields}\}$.  

We now observe that the  vacuum energy coming from particle physics enters the action via a term of the form $-2\Lambda \int d^4 x\sqrt{-\tilde g}$. This has no effect on the dynamics provided
\be
\frac{\delta}{\delta \phi_i}  \int d^4 x\sqrt{-\tilde g}=0 \label{cond} .
\ee
This is only possible when $\tilde g_{ab}$ is a composite field, for which $\sqrt{-\tilde g}$ is the integrand of a  topological invariant, and/or a total derivative. Note that our method is distinct from unimodular gravity in which the metric determinant is {\it constrained} to be unity \cite{unimodular}.

As an example, consider the case where the physical metric $\tilde g_{ab}$ is conformally related to the fundamental metric, $g_{ab}$, 
\be \label{conf}
\tilde g_{ab}=\Omega(\phi_i, \partial \phi_i, \del \del \phi_i, \ldots ) g_{ab}
\ee
where $\Omega>0$. Then $\sqrt{-\tilde g}=\Omega^2 \sqrt{-g}$ so one could achieve the desired property by taking, say, $\Omega^2=\frac{1}{\sqrt{-g}} \partial_b (\sqrt{-g} A^b)$. Later on we will present an example in which we take $\Omega^2$ to include   the Gauss-Bonnet invariant.  Alternatively one could consider disformally related metrics (see eg. \cite{disformal}), as long as they satisfy  the desired property (\ref{cond}). For the remainder of this paper, we will focus on the case where the fundamental and physical metrics are conformally related, as in (\ref{conf}). We also remind the reader once more that the $\phi_i$ include the 
fundamental metric $g_{ab}$, so $\Omega$ in general depends on same, as well as other fields.

While the condition (\ref{cond}) guarantees that the vacuum energy does not enter the dynamics, that is only half of the story. We need to propose kinetic terms for the fundamental fields, $\phi_i$.  Standard kinetic terms for each field will result in a modified theory of gravity. Such models are strongly constrained by both observation and theoretical considerations (see \cite{review} for an extensive review of modified gravity, including these issues). The strongest observational constraints come from solar system physics, where typically one must make use of either chameleon \cite{cham} or Vainshtein effects \cite{vainsh} in order to suppress the gravity modifications at these scales.

One way to ensure compatibility with solar system physics is to take the kinetic part of the action to be the Einstein-Hilbert term, built out of the physical metric, $\int d^4 x \sqrt{-\tilde g} R(\tilde g)$. It would then follow that any vacuum solution of General Relativity (with arbitrary cosmological constant) should also be a vacuum solution of this theory. Since the geometry of the solar system is well approximated by the Schwarzschild metric, this suggests that observational constraints can be easily met. However, there is a price to pay. In order for (\ref{cond}) to hold, it is clear that $\tilde g_{ab}$ must depend on derivatives of the fundamental fields, $\partial \phi_i$. Thus a kinetic term of the form $\sqrt{-\tilde g} R(\tilde g)$ will contain higher derivatives and may suffer from an Ostrogradski ghost \cite{ostro}.  

As we will see later on, ghosts {\it can} be avoided with a suitable choice of conformal factor $\Omega(\phi_i, \partial\phi_i, \del \del \phi_i,\ldots)$. Thus we consider a theory  of the form,
\be \label{action}
S[\phi_i; \Psi_n]=\frac{1}{16\pi G} \int d^4 x \sqrt{-\tilde g}  R (\tilde g)+S_m[\phi_i; \Psi_n]
\ee
where the matter action, $S_m$,  describes matter fields $\Psi_n$ minimally coupled to the composite metric $\tilde g_{ab}=\Omega(\phi_i, \partial\phi_i, \del \del \phi_i,\ldots) g_{ab}$.  We now compute the equations of motion. Variation of the action with respect to the fundamental fields, $\phi_i$ yields the following\footnote{To derive this equation, we first use  Chain Rule, $\frac{\delta S}{\delta \phi_i(x)}=\int d^4 y  \frac{\delta S}{\delta \tilde g_{ab}(y)}\frac{\delta \tilde g_{ab}(y)}{\delta \phi_i(x)}$ and then use the fact that $\frac{\delta \tilde g_{ab}(y)}{\delta \phi_i(x)}=\Omega \left( \delta^c_a \delta^d_b-\frac{1}{4} \tilde g_{ab} \tilde g^{cd}\right)\frac{\delta  g_{ab}(y)}{\delta \phi_i(x)} +\frac{1}{2\sqrt{-\tilde g}}\tilde g_{ab} \frac{\delta}{\delta \phi_i(x)}  \sqrt{-\tilde g(y)}$},
\be \label{phieq}
\frac{\delta S}{\delta \phi_i}=\sqrt{-\tilde g} \Omega \left(\tilde E^{ab}-\frac{1}{4} \tilde E \tilde g^{ab} \right)\frac{\del g_{ab}}{\del \phi_i}+\frac{1}{2} {\cal O}_i( \tilde E) =0
\ee
where 
\be
\tilde E^{ab}=\frac{1}{\sqrt{-\tilde g}} \frac{\delta S}{\delta \tilde g_{ab}}  =-\frac{1}{16\pi G} \left[ \tilde G_{ab}-8 \pi G \tilde T_{ab} \right]
\ee
and $\tilde E=\tilde E^a{}_a$.  Here, for the time being at least,  we are  raising and lowering indices with $\tilde g^{ab}$ and  $\tilde g_{ab}$ respectively.  Note that the first term in equation (\ref{phieq}) only appears in the $g_{ab}$ equation of motion, since 
$$
\frac{\del g_{ab}}{\del \phi_i}=\begin{cases} 1 & \phi_i=g_{ab} \\
0 & \phi_i=\text{scalar fields, vector fields, other tensor fields} \end{cases}
$$
The physical energy-momentum tensor is given by $\tilde T^{ab}=\frac{2}{\sqrt{-\tilde g}} \frac{\delta S_m}{\delta \tilde g_{ab}}$, while 
the linear operator ${\cal O}_i$  acts on scalars and is defined as
\begin{multline} 
{\cal O}_i (Q)=\int d^4 y Q(y) \frac{\delta }{\delta \phi_i(x)}\sqrt{-\tilde g(y)}
=Q(x) \frac{\del \sqrt{-\tilde g(x)}}{\del \phi_i(x)}-\frac{\del}{\del x^a} \left( Q(x)  \frac{\del \sqrt{-\tilde g(x)}}{\del \del_a \phi_i(x)}\right)\\
+\frac{\del^2}{\del x^a \del x^b} \left( Q(x)  \frac{\del \sqrt{-\tilde g(x)}}{\del \del_a \del_b \phi_i(x)}\right)+\ldots
\end{multline}
For a constant, $c$, it is clear that
\be
{\cal O}_i(c)=c  \frac{\delta }{\delta \phi_i(x)}\int d^4 y \sqrt{-\tilde g(y)}=0
\ee
by the condition (\ref{cond}). This is crucial in eliminating the vacuum energy from the dynamics. Indeed, it is easy to check that the constant vacuum energy contribution, $\sigma$, to the energy-momentum tensor, $\tilde T_{ab}=-\sigma \tilde g_{ab}+\ldots$ drops out of the equations of motion entirely, as expected. 

To demonstrate compatibility with solar system tests,  consider a vacuum solution to GR, of the form
\be
\tilde G_{ab}=-\tilde \Lambda \tilde g_{ab}, \qquad  \tilde T_{ab}=-\sigma \tilde g_{ab}
\ee
where $\tilde \Lambda$ is the vacuum curvature.  The equations of motion (\ref{phieq}) are now satisfied automatically. To see this, note that $\tilde E_{ab}=(\text{const}) \tilde g_{ab}$. It follows that $\tilde E$=constant, and so ${\cal O}_i(\tilde E)=0$. Thus we have proven our claim that any vacuum solution of GR (with arbitrary cosmological constant, $\tilde \Lambda$)  is also a vacuum solution in our theory. This applies to (de Sitter) Schwarzschild, which is  an excellent approximation for the geometry of the solar system. Note that the vacuum curvature, $\tilde \Lambda$ can be chosen empirically, and is completely independent of the vacuum energy, $\sigma$.

Actually, we can go even further. {\it Any}  solution of GR, vacuum or otherwise,  is also a solution to our theory, whatever the value of the vacuum curvature. As the vacuum energy drops out of the dynamics,  we are free to choose the vacuum curvature with a clean conscience.   Indeed, one can straightforwardly check that the field equations are satisfied by the choice, 
\be
\tilde G_{ab}=-\tilde \Lambda \tilde g_{ab}+ \tau_{ab}, \qquad  \tilde T_{ab}=-\sigma \tilde g_{ab}+\tau_{ab}
\ee
where $\tau_{ab}$ describes the matter excitations above the vacuum, $\sigma$ is the vacuum energy, and $\tilde \Lambda$ is the vacuum curvature. This follows from the fact that the equations of motion are linear in $\tilde E_{ab}$, with constant contributions dropping out completely. In particular, this means that the standard $\Lambda$CDM cosmology, with $\Lambda$ chosen empirically without any concern, is a perfectly good solution to our theory, and does not suffer from the same fine tuning issues as the corresponding solution in GR.

\section{Avoiding Ostrogradski ghosts: an example}
It remains to check whether or not our theory is ghost-free. To this end, we shall now consider vacuum fluctuations about an appropriate background. Let us start with a maximally symmetric vacuum,  with physical Riemann curvature, 
$ 
 \tilde R_{abc}{}^d=\frac{\tilde \Lambda}{3}(\tilde g_{ac} \delta^d_b-\tilde g_{bc} \delta_a^d)
$
and assume, for simplicity, that the background value of $\Omega=\bar \Omega=$ constant. It follows that the background curvature describing the fundamental field is given by $R_{abc}{}^d=\frac{ \Lambda}{3}( g_{ac} \delta^d_b- g_{bc} \delta_a^d)$, where $\Lambda=\bar \Omega \tilde \Lambda$. 

We write the fluctuations in the fundamental fields as
\be
\{\delta \phi_i\}=\{ h_{ab}, \ldots\}
\ee
where $h_{ab}=\delta g_{ab}$. Henceforth it is convenient to change convention and {\bf  raise and lower indices using the {\it un}tilded background metric $g^{ab}$ and $g_{ab}$.}
After a relatively lengthy calculation, we arrive at the following action describing the vacuum fluctuations to quadratic order,
\be 
\delta_2 S =\frac{\bar \Omega}{16 \pi G} \left[ \delta_2 S_{GR}[g] + \int d^4 x \sqrt{-g}\Delta {\cal L}\right] \label{d2S}
\ee
where $\delta_2 S_{GR}$ is the expansion to quadratic order (in $h_{ab}$)  of the standard Einstein-Hilbert action, with a cosmological constant, 
$$S_{GR}[g]=\int d^4 x \sqrt{-g} (R(g)-2\Lambda)$$
where $R(g)$ is the Ricci scalar built  from $g_{ab}$. In addition, we have a perturbative correction due to the modfiication of gravity given by
\be
\Delta {\cal L}  = 
\frac{1}{4} \frac{\delta  \Omega^2}{\bar \Omega^2}\left(2 \delta R(g)-\frac{3}{2} \frac{\nabla^a \nabla_a  \delta \Omega^2}{\bar \Omega^2}-2 \Lambda \frac{\delta \Omega^2}{\bar \Omega^2}  \right) 
\ee
where $\delta R(g)=\nabla_a\nabla_b (h^{ab}-h g^{ab})-\Lambda h$ is the linearised Ricci scalar (with metric connection), $h=h^a{}_a$ and $\nabla$ denotes the covariant derivative using the metric connection for $g_{ab}$. 

These expressions hold for any choice of $\Omega^2$ for which (\ref{cond}) holds.  Let us now consider the following special case
\be \label{choice}
\Omega^2=\frac{R_{GB}(\Xi, g)}{\mu^4} 
\ee
where  $\mu$ is a mass scale and
\be
R_{GB}(\Xi, g)=\frac{1}{4} \delta^{a_1 \cdots a_4}_{b1 \cdots b_4} R_{a_1 a_2}{}^{b_1 b_2} (\Xi, g) R_{a_3 a_4}{}^{b_3 b_4}(\Xi, g)
\ee
is the Gauss-Bonnet combination. Here $ \delta^{a_1 \cdots a_n}_{b_1 \cdots b_n}=n!  \delta^{a_1 \cdots a_n}_{[b_1 \cdots b_n]}$ is the generalised Kronecker delta symbol,  and $\Xi^c{}_{ab}$ is an {\it independent} torsion-free connection from which we  construct the corresponding Riemann tensor
 \ba
R_{ab}{}^{cd}(\Xi, g) &=& g^{c e} R_{abe}{}^d(\Xi) \nonumber \\
&=& g^{ce} (-2 \partial_{[a} \Xi^d{}_{b]e}+2 \Xi^f{}_{e[a} \Xi^d{}_{b]f} )
\ea
As is well known, $\sqrt{-g}R_{GB}(\Xi, g)$ is the Gauss-Bonnet invariant. It is, of course, topological in four dimensions, even when we take a Palatini variation for which $\Xi^c{}_{ab}$ is independent of $g_{ab}$. This guarantees that the property (\ref{cond}) is satisfied.

Let us assume that $\Xi^c{}_{ab}$ coincides with the metric connection $ \Gamma^c{}_{ab}=\frac{1}{2} g^{cd} (g_{da, b}+g_{db, a}-g_{ab, d})$ on the background, and we write the gauge invariant fluctuation as 
\be
\delta (\Xi^c{}_{ab}-  \Gamma^c{}_{ab})={\cal B}^c{}_{ab}
\ee
It follows that $\bar \Omega^2=\frac{8\Lambda^2}{3\mu^4 }$, and
\be
\frac{\delta \Omega^2}{\bar \Omega^2}=\frac{1}{2\Lambda} \left(\delta R(g)+\nabla_a {\cal X}^a \right)
\ee
where  ${\cal X}^a={\cal B}^{ab}{}_{b}- {\cal B}_b{}^{ba}$. Plugging this into the effective action (\ref{d2S}), we have that
\be \label{dL} 
\Delta {\cal L}  =  
\frac{1}{4} \psi \left(2 \delta R(g)-\frac{3}{2} \nabla^a \nabla_a  \psi -2 \Lambda \psi \right) \\
+\lambda \left( \delta R(g)+\nabla_a {\cal X}^a -2 \Lambda \psi\right)
\ee
where $\lambda$ is a Lagrange multiplier that fixes $\psi$ so that it coincides with $\frac{\delta \Omega^2}{\bar \Omega^2}$. 

Now, it is immediately obvious that the perturbative structure of our theory can only differ from GR in the scalar sector. This means that we have two well behaved propagating modes of spin 2, and none with spin 1. But what of the spin 0 modes? Are they pathological? Fortunately, for the choice of conformal factor (\ref{choice}), the spin 0 modes are also absent and there is no pathology. To see this we note that the precise form of  (\ref{dL})  suggests that the action (\ref{d2S}) can be rewritten as
 \be
\delta_2 S =\frac{\bar \Omega}{16 \pi G} \left[ \delta_2 S_{GR}[e^{\psi/2}g] \right.\\
\left.+ \int d^4 x \sqrt{-g}\lambda \left( \delta R(g)+\nabla_a {\cal X}^a -2 \Lambda \psi\right)\right] \label{d2S1}
\ee
where $\psi$ is understood to be small. That the effective action can ultimately be written like this was obvious given the form of full non-linear theory (\ref{action}). In any event, we now see that the ${\cal X}^a$ equation of motion yields 
$\del_a \lambda=0$. Assuming asymptotically vanishing boundary conditions, it follows that
$\lambda=0$.  This is enormously important, because integrating out the Lagrange multiplier, we find that the effective action is reduced to 
\be
\delta_2 S =\frac{\bar \Omega}{16 \pi G}  \delta_2 S_{GR}[e^{\psi/2}g]
\ee
But this is nothing more than the effective action for metric fluctuations of the Einstein-Hilbert action (with cosmological constant) on the maximally symmetric spacetime. The scalar $\psi$ simply acts to renormalise the scalar modes, but does not alter the fact that none of them propagate! This is shown explicitly in the appendix. The bottom line is that our theory has the same perturbative structure on maximally symmetric spaces as GR. In particular this means that we simply have two propagating tensor degrees of freedom and no ghost.  
\section{Summary}
We have proposed a novel way to clean up the cosmological constant problem.  By coupling matter to a composite metric, $\tilde g_{ab} (\phi, \partial \phi, \ldots)$, satisfying the property (\ref{cond}), we have been able to eliminate the troublesome vacuum energy from contributing to the dynamics of the system. Thus one ought to be able to choose the vacuum curvature to take on an empirical value, as dictated by observation, with a clean conscience.  This  is the take home message of this paper. The challenge now for model builders is to incorporate this idea into a viable model of gravity.  To this end we have proposed a model that exploits our neat idea, and at the same time ought to be ghost-free and compatible with solar system physics and cosmological tests.  This example contains a fundamental metric $g_{ab}$, and an independent torsion-free connection, $\Xi^c{}_{ab}$. It is described by the action (\ref{action}) with
\be
\tilde g_{ab}=\Omega g_{ab}, \qquad \Omega^2=\frac{R_{GB}(\Xi, g)}{\mu^4} 
\ee
and the matter fields, $\Psi_n$ are minimally coupled to the composite metric, $\tilde g_{ab}$.  Note that this choice of $\Omega^2$ is far from unique. One is free to add a whole host of terms  to its definition without introducing any unwanted pathologies.  This includes terms proportional to $R_{GB}(\Gamma, g)$,  and the Pontryagin term.  Indeed, all we need is for the conformal factor to contain some sort of ``auxiliary" field whose equation of motion constrains the Lagrange multiplier to vanish. If that is the case, the perturbative analysis goes through untroubled and there are no ghosts, at least not on maximally symmetric space. What about more general backgrounds? Whilst a detailed analysis of this is beyond the scope of this paper, we will say that we are optimistic that ghosts will remain absent. The point is that all we need is for the ``auxiliary" field to continue to constrain the Lagrange multiplier, allowing the rest of the perturbative analysis to mirror the case of General Relativity. Note that in the example presented here, the role of the  ``auxiliary" field  is played by the independent connection,  $\Xi^c{}_{ab}$.

As we have seen, any solution to GR, with arbitrary cosmological constant,  is a solution to our theory. However, it is possible that the reverse may  not be  true and a general  theory is expected to permit solutions that are not present in GR. Preliminary studies suggest that for the model given by equation (\ref{choice}), no further solutions exist, and the  extension to GR is fully encoded by  the arbitrariness in $\Lambda$. This suggests that the number of propagating degrees of freedom should be equivalent to GR. More general  models  may yield more exotic solutions containing interesting and potentially testable new features. Work is under way to  study the impact of these, beginning with cosmological solutions.  Even so, we emphasize, once again, that the main point of this paper is to propose the general idea, as given by (\ref{cond}), so we welcome any consistent model that attempts to incorporate this, either through a simple extension of our specific model, or by developing new models that embrace the spirit of the general idea.

\begin{acknowledgments} 
We would like to thank Ed Copeland, Nemanja Kaloper and Paul Saffin for useful discussions.
AP acknowledges  financial support from the Royal Society and IK from STFC. 
\end{acknowledgments}

\appendix
\section{Appendix} 
Starting from the effective action, $ \delta_2 S_{GR}[e^{\psi/2}g]$, let us briefly demonstrate that there are no propagating spin 0 modes on de Sitter space. Scalar perturbations of the de Sitter metric can be written as, 
\be
g_{ab} dx^a dx^b= -(1+\alpha)^2 dt^2 \\+e^{2Ht} e^{2\xi} \delta_{ij} (dx^i +\vec \nabla^i \beta dt) (dx^j +\vec \nabla^j \beta dt)
\ee
where, $H^2=\frac{\Lambda}{3}$ and, without loss of generality, we have chosen a gauge for which $\delta g_{ij}$ is a pure trace, thereby setting to zero the term of the form $(\vec \nabla_i \vec \nabla_j -\frac{1}{3} \delta_{ij} \vec \nabla^2 )\nu$.

It follows that the conformally related metric
\be
e^{\psi/2} g_{ab} dx^a dx^b= -(1+\tilde \alpha)^2 dt^2 \\+a^2(t) e^{2\tilde \xi} \delta_{ij} (dx^i +\vec \nabla^i \tilde \beta dt) (dx^j +\vec \nabla^j \tilde \beta dt)
\ee
where
\be
\tilde \alpha=\alpha+\frac{\psi}{4}, \qquad \tilde \xi=\xi+\frac{\psi}{4}, \qquad \tilde \beta=\beta
\ee
This is the renormalisation of the scalar modes alluded to in the main draft. It now follows that 
\begin{multline}
\delta_2 S_{GR}[e^{\psi/2}g]=\int dt d^3 x e^{3Ht}\{-6 \dot  {\tilde  \xi}^2-2e^{-2Ht}{\tilde  \xi} \vec \nabla^2 {\tilde  \xi} 
-6H^2 {\tilde \alpha}^2+12H{\tilde \alpha} \dot {\tilde  \xi} \\-4 e^{-2Ht} {\tilde \alpha} \vec \nabla^2 {\tilde  \xi} 
+4 e^{-2Ht}\dot  {\tilde  \xi} \vec \nabla^2 {\tilde \beta}-4H e^{-2Ht} {\tilde \alpha} \vec \nabla^2 {\tilde \beta}  \}
\end{multline}
The Hamiltonian and momentum constraints yield,
$
{\tilde \alpha}=\frac{\dot {\tilde  \xi}}{H}, ~ {\tilde \beta}=-\frac{{\tilde  \xi}}{H}$.
Integrating out ${\tilde \alpha}$ and ${\tilde \beta}$ (the lapse and shift), we find that the effective action describing the scalar perturbations vanishes completely. Thus there are no propagating spin 0 degrees of freedom.


\begin{thebibliography}{99}
\bibitem{sn}
  S.~Perlmutter {\it et al.} [ Supernova Cosmology Project Collaboration ],
  Astrophys.\ J.\  {\bf 517 } (1999)  565-586.
  [astro-ph/9812133].
\bibitem{wmap}
  D.~Larson {\it et al.},
  Astrophys.\ J.\ Suppl.\  {\bf 192} (2011) 16
  [arXiv:1001.4635 [astro-ph.CO]].

\bibitem{pauli}
  S.~E.~Rugh and H.~Zinkernagel,
  arXiv:hep-th/0012253.
  
\bibitem{nogo}
  S.~Weinberg,
  Rev.\ Mod.\ Phys.\  {\bf 61 } (1989)  1-23.
  
\bibitem{sled}
  C.~P.~Burgess,
  Annals Phys.\  {\bf 313 } (2004)  283-401.
  [hep-th/0402200].  
\bibitem{fab4}
   C.~Charmousis, E.~J.~Copeland, A.~Padilla and P.~M.~Saffin,
  arXiv:1106.2000 [hep-th].
   C.~Charmousis, E.~J.~Copeland, A.~Padilla and P.~M.~Saffin,
  arXiv:1112.4866 [hep-th].
  
\bibitem{klink}
  F.~R.~Klinkhamer, G.~E.~Volovik,
  JETP Lett.\  {\bf 91 } (2010)  259-265.
  [arXiv:0907.4887 [hep-th]].
  
\bibitem{bauer}
  F.~Bauer, J.~Sola, H.~Stefancic,
  JCAP {\bf 1012 } (2010)  029.
  [arXiv:1006.3944 [hep-th]].
  
\bibitem{Zonetti}
  J.~Govaerts and S.~Zonetti,
  Class.\ Quant.\ Grav.\  {\bf 28} (2011) 185001
  [arXiv:1102.4957 [hep-th]].
  
\bibitem{Shaw}
  D.~J.~Shaw and J.~D.~Barrow,
  Phys.\ Rev.\ D {\bf 83} (2011) 043518
  [arXiv:1010.4262 [gr-qc]].
  J.~D.~Barrow and D.~J.~Shaw,
  Gen.\ Rel.\ Grav.\  {\bf 43} (2011) 2555
   [Int.\ J.\ Mod.\ Phys.\ D {\bf 20} (2011) 2875]
  [arXiv:1105.3105 [gr-qc]].
  
\bibitem{naresh}
  N.~Dadhich,
  Int.\ J.\ Mod.\ Phys.\ D {\bf 20} (2011) 2739
  [arXiv:1105.3396 [gr-qc]].
  N.~Dadhich,
  Pramana {\bf 77} (2011) 433
  [arXiv:1006.1552 [gr-qc]].
  
\bibitem{ww}
  S.~Weinberg and E.~Witten,
  Phys.\ Lett.\ B {\bf 96} (1980) 59.
  
\bibitem{unimodular}
  J.~L.~Anderson and D.~Finkelstein,
  Am.\ J.\ Phys.\  {\bf 39} (1971) 901.
  
\bibitem{disformal}
  N.~Kaloper,
  Phys.\ Lett.\  B {\bf 583} (2004) 1
  [arXiv:hep-ph/0312002].
  
\bibitem{review}
  T.~Clifton, P.~G.~Ferreira, A.~Padilla, C.~Skordis,
  [arXiv:1106.2476 [astro-ph.CO]].
  

  
\bibitem{cham}
  J.~Khoury, A.~Weltman,
  Phys.\ Rev.\ Lett.\  {\bf 93 } (2004)  171104.
  [astro-ph/0309300].
  
\bibitem{vainsh}
  A.~I.~Vainshtein,
  Phys.\ Lett.\  {\bf B39 } (1972)  393-394.
  N.~Kaloper, A.~Padilla, N.~Tanahashi,
  [arXiv:1106.4827 [hep-th]].

\bibitem{ostro}
M. Ostrogradski, 
Memoires de lAcademie Imperiale des Science de Saint-Petersbourg, 4:385, 
1850. 


\end{thebibliography}
\end{document}